\documentclass[aps,prd,twocolumn,superscriptaddress]{revtex4}
\pdfoutput=1

\usepackage{amssymb}
\usepackage{amsmath}
\usepackage{graphicx}

\newcommand{\be}{\begin{equation}}
\newcommand{\ee}{\end{equation}}

\newcommand{\Eq}[1]{eq.~(\ref{#1})}

\makeatletter
\def\simgt{\mathrel{\lower2.5pt\vbox{\lineskip=0pt\baselineskip=0pt
           \hbox{$>$}\hbox{$\sim$}}}}
\def\simlt{\mathrel{\lower2.5pt\vbox{\lineskip=0pt\baselineskip=0pt
           \hbox{$<$}\hbox{$\sim$}}}}
\makeatother

\begin{document}

\title{Collective Quartics and Dangerous Singlets in Little Higgs}

\author{Martin Schmaltz}
\affiliation{Physics Department, Boston University, Boston, MA 02215}
\affiliation{Berkeley Center for Theoretical Physics,
  University of California, Berkeley, CA 94720}
\affiliation{Theoretical Physics Group, Lawrence Berkeley
  National Laboratory, Berkeley, CA 94720}
  
\author{Jesse Thaler}
\affiliation{Berkeley Center for Theoretical Physics,
  University of California, Berkeley, CA 94720}
\affiliation{Theoretical Physics Group, Lawrence Berkeley
  National Laboratory, Berkeley, CA 94720}

\begin{abstract}
Any extension of the standard model that aims to describe TeV-scale physics without fine-tuning must have a radiatively-stable Higgs potential.  In little Higgs theories, radiative stability is achieved through so-called collective symmetry breaking.  In this letter, we focus on the necessary conditions for a little Higgs to have a collective Higgs quartic coupling.  In one-Higgs doublet models, a collective quartic requires an electroweak triplet scalar.  In two-Higgs doublet models, a collective quartic requires a triplet or singlet scalar.  As a corollary of this study, we show that some little Higgs theories have dangerous singlets, a pathology where collective symmetry breaking does not suppress quadratically-divergent corrections to the Higgs mass.
\end{abstract}

\maketitle

\section{Introduction}

The standard model Higgs mass gets quadratically-divergent radiative corrections from electroweak gauge interactions, the top quark Yukawa coupling, and the Higgs quartic interaction. These radiative corrections become large and require fine-tuning of the Higgs potential when one pushes the range of validity of the theory above the TeV scale.  Thus, any model that is designed to describe physics at LHC energies without fine-tuning must incorporate additional structures in the gauge, top, and Higgs sectors to remove these quadratic divergences.

Little Higgs theories \cite{ArkaniHamed:2001nc,ArkaniHamed:2002qx,ArkaniHamed:2002qy,Schmaltz:2005ky,Perelstein:2005ka} avoid quadratic divergences through collective symmetry breaking.  In the quartic sector, for example, the Higgs quartic coupling is introduced through two operators, both of which individually preserve enough symmetries to forbid radiative corrections to the Higgs mass, but collectively generate the desired Higgs potential. While this recipe sounds straightforward, there are known examples in the literature \cite{Schmaltz:2004de,Cheng:2006ht} where collectively generating gauge/fermion couplings is possible, but implementing a collective quartic appears to be impossible.  

This difficulty of constructing little Higgs quartics motivates us to examine the structure of quartic couplings with collective symmetry breaking in a systematic way.  Our main result is that a successful collective quartic requires additional scalars with specific electroweak quantum numbers.  In particular, the quartic of a one-Higgs doublet model requires (complex or real) $SU(2)_L$ triplets, while the quartic of a two-Higgs doublet model can be constructed with either triplets or singlets, as long as the singlet carries some non-trivial global charge. 

Moreover, we find that real singlet scalars pose a potential danger to Higgs mass stability in little Higgs models.
The problem arises when the shift symmetry which would na\"{\i}vely protect the Higgs boson mass
\be
h \rightarrow h + \epsilon + \cdots
\ee
is accompanied by shifts acting on a real singlet $\eta$
\be
\eta \rightarrow \eta \mp \frac{\epsilon^\dagger h + h^\dagger \epsilon}{f} + \cdots ,
\ee
where $f$ is the decay constant of some non-linear sigma model.  In this case, the operators
\be
\label{eq:badmass}
\mathcal{L} = M^3 \left (\eta \pm \frac{h^\dagger h}{f} + \cdots \right)
\ee
are invariant under the combined shift symmetries and contain Higgs mass terms.  This is the problem of dangerous singlets in little Higgs theories. To ensure that operators like \Eq{eq:badmass} are not radiatively generated, one must preserve additional symmetries acting on $\eta$.

These results clarify the known quartic mechanisms in the little Higgs literature.  The $SU(5)/SO(5)$ littlest Higgs \cite{ArkaniHamed:2002qy} is an example of a one-Higgs doublet model with an additional complex triplet.  The $SU(6)/Sp(6)$ antisymmetric condensate model \cite{Low:2002ws} is an example of a two-Higgs doublet model with an additional complex singlet.  Our arguments explain why any attempt in one-Higgs doublet models to build quartics with only additional singlets is destined to fail.  As cautionary examples of dangerous singlets, the $SO(9)/(SO(5)\times SO(4))$ \cite{Chang:2003zn} and $SU(9)/SU(8)$ \cite{Skiba:2003yf} models both have unacceptable quadratically-divergent contributions to the Higgs mass.   

The difficulties with constructing quartics are not limited to little Higgs theories, and similar issues appear in certain extra-dimensional models with bulk gauge/fermion fields and brane-localized symmetry breaking \cite{Contino:2003ve,Agashe:2004rs}.  While extra-dimensional locality guarantees collective symmetry breaking in the gauge and fermion sectors, locality does not imply collective symmetry breaking in the quartic sector.  In models like \cite{Contino:2003ve,Agashe:2004rs}, a quartic coupling can be generated through fine-tuning, but to construct a naturally large quartic coupling, one needs to introduce collective structures (as also suggested in \cite{Contino:2003ve}).  The results of this letter pertain to these natural quartic mechanisms.

In the next section, we classify all possible little Higgs quartics in one- and two-Higgs doublet models according to $SU(2)_L$ transformation properties and show why quartics cannot arise from singlet scalars.  In section~\ref{sec:dangeroussinglet}, we discuss the problem of dangerous singlets.  We conclude with some lessons for little Higgs model building.   An example model is presented in the appendix.

\section{Collective Quartics}
\label{sec:collectivequartics}

How does one construct a Higgs quartic that does not radiatively generate a quadratically-divergent Higgs mass?  In little Higgs theories, one finds a set of operators that each preserve different shift symmetries acting on the Higgs doublet, but collectively break all the symmetries that protect the Higgs potential \cite{ArkaniHamed:2001nc,ArkaniHamed:2002qx,ArkaniHamed:2002qy,Schmaltz:2005ky,Perelstein:2005ka}.  Since the quadratically-divergent diagrams only involve one operator at a time, the shift symmetries are sufficient to protect the Higgs mass parameter.

Concretely, in a non-linear sigma model where the Higgs is a pseudo-Nambu-Goldstone boson (PNGB), one na\"{\i}vely expects the shift symmetry
\be
\label{eq:higgsshift}
h \rightarrow h + \epsilon + \cdots
\ee
to forbid any potential for the Higgs.  But if there are additional PNGBs $\phi$ with compensating shifts
\be
\label{eq:phishift}
\phi \rightarrow \phi \mp \frac{h \epsilon + \epsilon h}{f} + \cdots,
\ee 
then the two operators
\be
\label{eq:quarticoperator}
V \sim \lambda_1 f^2  \left |\phi + \frac{h^2}{f}  + \cdots \right|^2 + \lambda_2 f^2  \left |\phi - \frac{h^2}{f}  + \cdots \right|^2
\ee
each preserve one of the Higgs shift symmetries from \Eq{eq:phishift}.  Taken alone, neither $\lambda_i$ term would give a physical Higgs quartic since each individual quartic could be removed by a $\phi_\pm \equiv \phi \pm h^2/f + \cdots$ field redefinition. Collectively, though, the two operators yield a Higgs quartic after $\phi$ is integrated out:
\be
V \sim \frac{4 \lambda_1 \lambda_2}{\lambda_1 + \lambda_2} h^4 + \cdots.
\ee
This is the form of all little Higgs quartics.  A small Higgs mass term is generated radiatively from \Eq{eq:quarticoperator}, and the resulting potential allows for a parametric separation between the electroweak vev $v$ and the decay constant $f$.  

At this point, we have not specified the quantum numbers of the scalar $\phi$, which is equivalent to specifying the quantum numbers of $h^2$.  The possible $SU(2)_L$ representations for $h^2$ are determined by
\be
\mathbf{2} \otimes \mathbf{2} = \mathbf{3}_S \otimes \mathbf{1}_A, \qquad \mathbf{2} \otimes \mathbf{\overline{2}} = \mathbf{3} \otimes \mathbf{1},
\ee
where the $S$/$A$ subscript refers to the representation being symmetric/antisymmetric under the interchange of the two doublets.  This classification holds regardless of the number of Higgs fields.

In a one-Higgs doublet model, the $\mathbf{1}_A$ representation vanishes, and $\phi$ can be a complex triplet, a real triplet, or a real singlet:
\begin{align}
h^i h^j &\rightarrow \phi^{ij} & (\mathbf{3}_S), \\
h^\dagger \tau^a h &\rightarrow \phi^{a}& (\mathbf{3}),\\
h^\dagger h &\rightarrow \eta& (\mathbf{1}),
\end{align}
where $\tau^a$ are the Pauli matrices, and we use the notation $\eta$ to refer to a real singlet that carries no other charges.   If $\phi$ is a real or complex $SU(2)_L$ triplet, then \Eq{eq:quarticoperator} gives rise to a tree-level quartic coupling yet protects the Higgs mass.  A complex $\phi$ triplet is used in the $SU(5)/SO(5)$ littlest Higgs \cite{ArkaniHamed:2002qy}, and a real $\phi$ triplet is present in the $SO(9)/(SO(5)\times SO(4))$ construction \cite{Chang:2003zn} (though this latter model has a pathology that will be understood in the next section).

However, if $\phi$ is a real singlet $\eta$, then explicit computation shows that \Eq{eq:quarticoperator} generates a quadratically-divergent $\eta$ tadpole and Higgs mass at one-loop!  (For an example, see the appendix.)
\be
\label{eq:phimass}
\frac{\lambda_1 f \Lambda^2}{16\pi^2} \left (\!\eta + \frac{h^\dagger h}{f}  + \cdots\! \right) - \frac{\lambda_2 f \Lambda^2}{16\pi^2} \left (\!\eta - \frac{h^\dagger h}{f}  + \cdots \!\right)
\ee
Note the sign difference between the two terms, which means that the Higgs mass term cannot be forbidden by $T$-parity \cite{ArkaniHamed:2002qy,Cheng:2003ju,Cheng:2004yc} with $\lambda_1 = \lambda_2$, and a parity that enforces $\lambda_1 = - \lambda_2$ would imply no Higgs quartic coupling in the first place.  Therefore, there is no viable one-Higgs doublet little Higgs model where a collective quartic involves a real singlet $\eta$.  In particular, this explains why it is impossible to add a collective quartic coupling to the simplest little Higgs \cite{Schmaltz:2004de} without extending the Higgs sector \cite{Kaplan:2003uc}.

The reason for this pathology is that the shift symmetry alone does not forbid a tadpole for $\eta$.  If $\eta$ had non-trivial quantum numbers (such as being an $SU(2)_L$ triplet), then these extra symmetries would forbid the $\eta$ tadpole.  Famously, the singlet $h^\dagger h$ cannot be charged under \emph{any} symmetry (except a shift symmetry), and the same holds for the singlet $\eta$.   To illustrate this pathology further, we construct an explicit singlet $\eta$ model which realizes the full non-linear shift symmetries in appendix~\ref{sec:realsingletmodel1}.

In a two-Higgs doublet model, one can have quartics constructed not only with $SU(2)_L$ triplets but also with singlets.  Choosing conventions where $h_1$ and $h_2$ have the same hypercharge, $\phi$ can \emph{a priori} be a complex singlet with or without hypercharge:
\begin{align}
h_1^i h_2^j \epsilon_{ij} &\rightarrow \phi & (\mathbf{1}_A),\\
h_1^\dagger h_2 &\rightarrow \phi & (\mathbf{1}).
\end{align}
Note however that the quartic constructed from the hypercharge carrying singlet $|h_1^i h_2^j \epsilon_{ij}|^2$ is unsatisfactory because it vanishes when the $h_1$ and $h_2$ vevs are aligned to preserve electric charge.  A hypercharge neutral complex $\phi$ is used in the $SU(6)/Sp(6)$ antisymmetric condensate model \cite{Low:2002ws}.  

In addition, $\phi$ can even be a real singlet as long as it has an extra $\mathbf{Z}_2$ symmetry:
\begin{align}
\mathrm{Re}[h_1^\dagger h_2] &\rightarrow \phi & (\mathbf{1}).
\end{align}
In this case, the symmetry
\be
\phi \rightarrow -\phi, \qquad h_1 \rightarrow - h_1, \qquad h_2 \rightarrow h_2
\ee
is sufficient to forbid the $\phi$ tadpole.

In summary, to construct a collective quartic coupling in little Higgs theories, one must have an additional scalar $\phi$ that both shifts according to \Eq{eq:phishift} and has additional symmetries that forbid a $\phi$ tadpole.

\section{Dangerous Singlets}
\label{sec:dangeroussinglet}

We argued that a satisfactory Higgs quartic coupling cannot be constructed using a singlet scalar $\eta$ that carries no other charges.  A corollary to this argument is that whenever a little Higgs model has a singlet scalar $\eta$, one must make sure not to introduce large operators that preserve only the shift symmetry 
\be
\label{eq:etashift}
h \rightarrow h + \epsilon, \qquad \eta \rightarrow \eta \mp \frac{h^\dagger \epsilon + \epsilon^\dagger h}{f},
\ee
under which $(f \eta \mp h^\dagger h)$ is invariant.  In order to prevent dangerous $\eta$ tadpoles (and corresponding $h^\dagger h$ mass terms) from being generated radiatively, additional symmetries are required under which $\eta$ transforms either linearly or non-linearly.

Note that the Lagrangian term $M^2(f\eta + h^\dagger h)$ by itself does not contribute to the Higgs mass, since a field redefinition on $\eta$ can remove the apparent Higgs mass term.  Only when both terms $M^2(f\eta \pm h^\dagger h)$ are present does the Higgs mass become physical.  This is in precise analogy with the quartic operators in \Eq{eq:quarticoperator}, where the $\lambda_1$ term only becomes a physical Higgs quartic coupling in conjunction with the $\lambda_2$ term.  Unfortunately both types of $\eta$ tadpoles do typically appear in collective symmetry breaking scenarios.

The dangerous singlet pathology is present in the $SO(9)/(SO(5)\times SO(4))$ construction \cite{Chang:2003zn}.  This model is designed to preserve custodial $SU(2)$, therefore the PNGBs are classified in terms of $SO(4)=SU(2)_L\times SU(2)_R$ representations. Under this $SO(4)$ the Higgs transforms as a $\mathbf{4}$, whereas the PNGBs which play the role of $\phi$ transform in the symmetric product of two $\mathbf{4}$'s. This symmetric product is reducible and it contains a 9-dimensional symmetric tensor as well as a dangerous singlet $\phi^0$.   The radiative potential generated by the gauge interactions contains a good quartic involving the symmetric tensor but also a bad quartic involving the singlet
\be
V = \lambda_\pm f  
\left |\phi^0 \pm \frac{h^\dagger h}{f}  + \cdots \right|^2 ,
\ee
which leads to a quadratically-divergent Higgs mass at one loop.

The pathology is also present in the $SU(9)/SU(8)$ construction \cite{Skiba:2003yf}, albeit in a more subtle way.  In this model, the operators that generate the quartic coupling also radiatively induce a quadratically-divergent tadpole for a field $s_2^I$.  The $s_2^I$ tadpole has a similar form as \Eq{eq:phimass}, where $h^\dagger h$ is replaced by $h_1^\dagger h_1$. Thus the dangerous singlet $s_2^I$ brings with it a quadratically divergent mass term for one of the two Higgs doublets in the model. 
The authors of Ref.~\cite{Skiba:2003yf} do consider ways to stabilize the $s_2^I$ vev, but unfortunately, no matter how $s_2^I$ is stabilized, the mass term for $h_1$ has already been generated.  For example, there is a $T$-symmetric limit which is the analog of setting $\lambda_1 = \lambda_2$ in \Eq{eq:phimass}, and while the $s_2^I$ tadpole vanishes in this limit, the Higgs mass term remains.

In general, a simple way to avoid a dangerous singlet is to maintain the full non-linear shift symmetry on $\eta$. In that case, $\eta$ is an exact NGB of a spontaneously broken $U(1)$ symmetry and the $\eta$ tadpole is forbidden. To avoid the associated massless particle, one could then softly break the symmetry or else gauge the $U(1)$ to eat the NGB. Unfortunately, in the examples considered above, the symmetry is broken by the gauge interactions so that the dangerous singlet cannot be avoided in this way. 

\section{Little Lessons}

In this letter, we have outlined the minimal requirements to get a collective quartic coupling in little Higgs theories.  One must introduce extra scalars $\phi$ that not only shift according to \Eq{eq:phishift},
but also carry additional charges that forbid a $\phi$ tadpole.  Moreover, one must make sure that there are no dangerous singlet scalars $\eta$ that can shift collectively.

The fact that one-Higgs doublet models \emph{require} a Higgs triplet may have interesting LHC implications.  The little M-theory construction \cite{Cheng:2006ht} was introduced as a phenomenological little Higgs model for LHC studies.  However, the $Sp(4)/SU(2)$ version not only does not have a Higgs quartic, but \emph{cannot} have a quartic without an additional Higgs field or an $SU(2)_L$ triplet.  Therefore, the little M-theory spectrum may not be representative of the LHC-accessible field content of a complete little Higgs theory.

Another observation about triplet scalars is that they usually get vevs after electroweak symmetry breaking, which induces a large correction to the $T$ parameter.  This suggests that generically, one-Higgs doublet models need $T$-parity \cite{Cheng:2003ju,Cheng:2004yc} in order to forbid a triplet tadpole.  Two-Higgs doublet models do not need to have $SU(2)_L$ triplet fields and are therefore less constrained.

We find it curious that the literature contains no one-Higgs doublet model without dangerous singlets where a collective quartic is constructed with a real $SU(2)_L$ triplet.  We challenge the little Higgs community to construct such a model or prove a no-go theorem.

Finally, given the quartic coupling constraints, our result suggests a new strategy for little Higgs model building.  In the past, little Higgs models typically started by collectively coupling the Higgs to gauge fields, then collectively adding the top Yukawa, and then introducing a Higgs quartic.   A new approach to solving the little hierarchy problem would be to start with a collective Higgs quartic at the outset, and then collectively add gauge and Yukawa couplings using known mechanisms in the literature \cite{Contino:2003ve,Thaler:2005kr}.  Such quartic-motivated models may have different symmetries and spectra compared to existing little Higgs theories and could be less fine-tuned.

\acknowledgments{
We benefitted from conversations with Spencer Chang, Andrew Cohen, Zoltan Ligeti, Witold Skiba, Daniel Stolarski, John Terning, and Jay Wacker.   M.S. thanks the Berkeley CTP and the LBNL particle theory group for their hospitality. M.S. is supported by DE-FG02-91ER-40676 and DE-FG02-01ER-40676.  J.T. is supported by the Miller Institute for Basic Research in Science.}

\appendix

\section{An Inviable Little Higgs with a Real Singlet and One Higgs Doublet}
\label{sec:realsingletmodel1}

In this appendix, we construct a little Higgs model with a collective quartic coupling using a real singlet $\eta$.  We will not put in fermion/gauge partners, because that can easily be done in an extra-dimensional picture.  We will see that the resulting quartic coupling radiatively generates a quadratically-divergent Higgs mass, and therefore this model is inviable.  The main result beyond the arguments from section~\ref{sec:collectivequartics} is that this model includes the full non-linear PNGB structure. 

To obtain the correct number of PNGBs, consider the symmetry breaking pattern $SO(6)/SO(5)$, which follows from a $\mathbf{6}$ of $SO(6)$ getting a vev. The PNGBs may be parametrized in terms of a linear sigma field $\Phi$ by writing
\be
\Phi = e^{i \Pi/ f} \left( \begin{array}{c} 0 \\ 0 \\ f \end{array}   \right), \qquad \Pi =  \frac{i}{\sqrt{2}} \left(\begin{array}{ccc} 0 & 0 & h \\ 0 & 0 & \eta \\ -h^T & -\eta & 0 \end{array}  \right),
\ee
where $\Phi$ is the $\mathbf{6}$ of $SO(6)$, $h$ is a real $\mathbf{4}$ of $SO(4)$ which contains the electroweak $SU(2)_L$ under which $h$ is a complex doublet, and $\eta$ is a real singlet.

Next, we need two operators that preserve different shift symmetries acting on the Higgs
\be
\label{eq:appendixquartic}
\mathcal{L} = \lambda_1 (\Phi^\dagger P_1 \Phi)^2 +  \lambda_2 (\Phi^\dagger P_2 \Phi)^2 ,
\ee
where
\be
P_1 = \left(\begin{array}{ccc} 1 & 0 & 0 \\ 0 & 0 & 1 \\ 0 & 1 & 0 \end{array}  \right), \qquad P_2 = \left(\begin{array}{ccc} 1 & 0 & 0 \\ 0 & 0 & -1 \\ 0 & -1 & 0 \end{array}  \right).
\ee
Taken alone, these two operators preserve two different $SO(5)$ symmetries, which can be seen explicitly by diagonalizing the $P_i$.   Both $SO(5)$ symmetries are spontaneously broken by the $\Phi$ vev, thus each operator alone leaves the Higgs as an exact NGB.  Together, the two operators only preserve an $SO(4)$ symmetry, which allows a quartic coupling of the same form as \Eq{eq:quarticoperator}.  This is the essence of collective breaking.  

However,
\be
\label{eq:so6singletmasses}
\mathcal{L} = m_1^2 \text{Tr}\left[{P_1}\right] \Phi^\dagger P_1 \Phi +  m_2^2 \text{Tr}\left[{P_2}\right]  \Phi^\dagger P_2 \Phi
\ee
is not forbidden by any symmetry, and is in fact radiatively generated with a quadratic divergence:
\be
m_1^2 \simeq \lambda_1 \frac{\Lambda^2}{16\pi^2}, \qquad m_2^2 \simeq \lambda_2 \frac{\Lambda^2}{16\pi^2}.
\ee
When expanded out, \Eq{eq:so6singletmasses} contains a tadpole for $\eta$ and a Higgs mass term, just as \Eq{eq:phimass}.  Note that $T$-parity ($\lambda_1 = \lambda_2$) does not help, and reverse $T$-parity ($\lambda_1 = -\lambda_2$) implies a vanishing quartic.  Also, the $\text{Tr}\left[{P_i}\right]$ terms make clear that a spurion symmetry $P_i \rightarrow -P_i$ does not forbid \Eq{eq:so6singletmasses}.  

\begin{figure}[h]
\begin{center}
\includegraphics[scale=0.25]{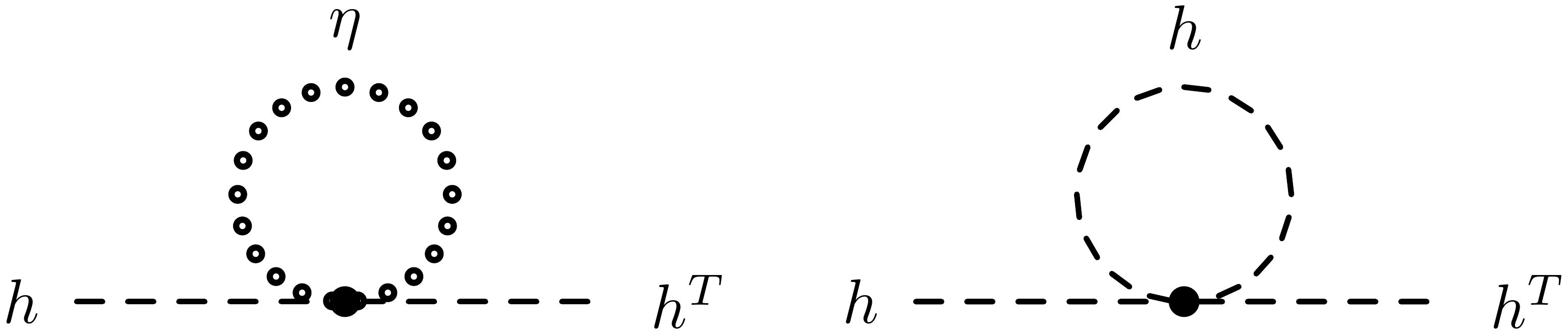}
\end{center}
\caption{The two quadratically divergent diagrams that contribute to the Higgs boson mass.
\label{fig:seagull}}
\end{figure}

To see the non-cancellation of the quadratic divergence in the Higgs mass diagrammatically, consider the $\eta$-loop and a Higgs loop as shown in figure~\ref{fig:seagull}. In a theory with a proper collective quartic, their contributions would be required to cancel by a symmetry. Here there is no such symmetry, and explicit computation shows that the quadratic divergence in the first diagram is
proportional to $-(\lambda_1+\lambda_2)$ whereas the second is proportional to $+3(\lambda_1+\lambda_2)$ so that they do not cancel.

\end{document}